\documentclass[journal=jacsat,manuscript=article]{achemso}
\usepackage{chemformula} 
\usepackage[T1]{fontenc} 
\usepackage{graphicx}
\usepackage{dcolumn}
\usepackage{bm}
\usepackage{xfrac}%
\usepackage{color}
\usepackage{tabularx}%
\usepackage{hyperref}
\usepackage{xcolor,colortbl}
\usepackage{multirow}

\newcommand{\RCL}{$\alpha$-RuCl$_{3}$}

\newcommand{\CIO}{Cu$_2$IrO$_3$}
\newcommand{\NIO}{Na$_2$IrO$_3$}
\newcommand{\LIO}{$\alpha$-Li$_2$IrO$_3$}
\newcommand{\LRO}{Li$_2$RhO$_3$}

\newcommand{\ALIO}{Ag$_{3}$LiIr$_2$O$_6$}
\newcommand{\ALRO}{Ag$_{3}$LiRh$_2$O$_6$}

\newcommand{\HLIO}{H$_{3}$LiIr$_2$O$_6$}
\newcommand{\CLIO}{Cu$_{3}$LiIr$_2$O$_6$}
\newcommand{\CNIO}{Cu$_{3}$NaIr$_2$O$_6$}

\newcommand{\tg}{$t_{\text {2g}}$}
\newcommand{\eg}{$e_{\text {g}}$}

\newcommand{\eeg}{$e'_{\text {g}}$}
\newcommand{\jeff}{$J_{\text {eff}}$}

\author{Faranak~Bahrami}
\affiliation{Department of Physics, Boston College, Chestnut Hill, MA 02467, USA}
\author{Xiaodong~Hu}
\affiliation{Department of Physics, Boston College, Chestnut Hill, MA 02467, USA}
\author{Yonghua~Du}
\affiliation{National Synchrotron Light Source II, Brookhaven National Laboratory, Upton, NY 11973, USA}
\author{Oleg~I.~Lebedev}
\affiliation{Laboratoire CRISMAT, ENSICAEN-CNRS UMR6508, 14050 Caen, France}
\author{Chennan~Wang}
\affiliation{Laboratory for Muon Spin Spectroscopy (LMU), Paul Scherrer Institute (PSI), CH-5232 Villigen, Switzerland}
\author{Hubertus~Luetkens}
\affiliation{Laboratory for Muon Spin Spectroscopy (LMU), Paul Scherrer Institute (PSI), CH-5232 Villigen, Switzerland}
\author{Gilberto~Fabbris}
\affiliation{Advanced Photon Source, Argonne National Laboratory, Argonne, IL 60439, USA}
\author{Michael~J.~Graf}
\affiliation{Department of Physics, Boston College, Chestnut Hill, MA 02467, USA}
\author{Daniel~Haskel}
\affiliation{Advanced Photon Source, Argonne National Laboratory, Argonne, IL 60439, USA}
\author{Ying~Ran}
\affiliation{Department of Physics, Boston College, Chestnut Hill, MA 02467, USA}
\author{Fazel~Tafti}
\affiliation{Department of Physics, Boston College, Chestnut Hill, MA 02467, USA}
\email{fazel.tafti@bc.edu}
\title{First demonstration of tuning between the Kitaev and Ising limits in a honeycomb lattice}

\begin{document}

\pagebreak
\begin{abstract}
Recent observations of novel spin-orbit coupled states have generated tremendous interest in $4d/5d$ transition metal systems.
A prime example is the $J_{\text{eff}}=\frac{1}{2}$ state in iridate materials and \RCL\ that drives Kitaev interactions. 
Here, by tuning the competition between spin-orbit interaction ($\lambda_{\text{SOC}}$) and trigonal crystal field splitting ($\Delta_\text{T}$), we restructure the spin-orbital wave functions into a novel $\mu=\frac{1}{2}$ state that drives Ising interactions.
This is done via a topochemical reaction that converts \LRO\ to \ALRO, leading to an enhanced trigonal distortion and a diminished spin-orbit coupling in the latter compound.
Using perturbation theory, we present an explicit expression for the new $\mu=\frac{1}{2}$ state in the limit $\Delta_\text{T}\gg \lambda_{\text{SOC}}$ realized in \ALRO, different from the conventional $J_\text{eff}=\frac{1}{2}$ state in the limit $\lambda_{\text{SOC}}\gg \Delta_\text{T}$ realized in \LRO. 
The change of ground state is followed by a dramatic change of magnetism from a 6~K spin-glass in \LRO\ to a 94~K antiferromagnet in \ALRO.
These results open a pathway for tuning materials between the two limits and creating a rich magnetic phase diagram. 
\end{abstract}

\pagebreak

\section*{Introduction}
An exotic quantum state in condensed matter physics is the $J_{\text{eff}}=\frac{1}{2}$ state in honeycomb iridate materials that leads to the Kitaev exchange interaction~\cite{takagi_concept_2019,knolle_field_2019,chaloupka_kitaev-heisenberg_2010,jackeli_mott_2009,kim_novel_2008,kitaev_anyons_2006}.
The $J_{\text{eff}}=\frac{1}{2}$ state is a product of strong spin-orbit coupling (SOC) in heavy Ir$^{4+}$ ions that splits the \tg\ manifold into a $J_{\text{eff}}=\frac{3}{2}$ quartet and a $J_{\text{eff}}=\frac{1}{2}$ doublet.
With five electrons in the $5d^5$ configuration, iridates have one electron in the spin-orbital $J_{\text{eff}}=\frac{1}{2}$ state that satisfies the prerequisites of the Kitaev interaction in a honeycomb lattice as shown by earlier studies~\cite{takagi_concept_2019,knolle_field_2019,chaloupka_kitaev-heisenberg_2010,jackeli_mott_2009,kim_novel_2008}.
In this article, we introduce a new spin-orbital state, $\mu=\frac{1}{2}$, which we have engineered by tuning the interplay between two energy scales: the SOC ($\lambda_\text{SOC}$) and the trigonal crystal field splitting ($\Delta_\text{T}$).
The $\mu=\frac{1}{2}$ state drives Ising instead of Kitaev interactions.
Although the Ising limit has been discussed in several theoretical studies~\cite{khaliullin_orbital_2005,bhattacharjee_spinorbital_2012,haraguchi_strong_2020,joy_magnetism_1992}, a transition between the Kitaev and Ising limits has not been demonstrated until now.
It has been theoretically predicted that the Kitaev limit in \NIO\ can be tuned to an Ising limit under uniaxial physical pressure~\cite{bhattacharjee_spinorbital_2012}, but the required pressure has not been achieved. 
The Ising limit is relevant to \emph{M}PS$_3$ (\emph{M} = Mn, Fe, Ni) compounds~\cite{joy_magnetism_1992}, however, a transition from the Ising to Kitaev limit has not been discussed in those materials, even at a theoretical level.
This work presents the first observation of a transition between the Kitaev and Ising limits in the same material family.

\begin{figure*}[ht]
 \centering
 \includegraphics[width=\textwidth]{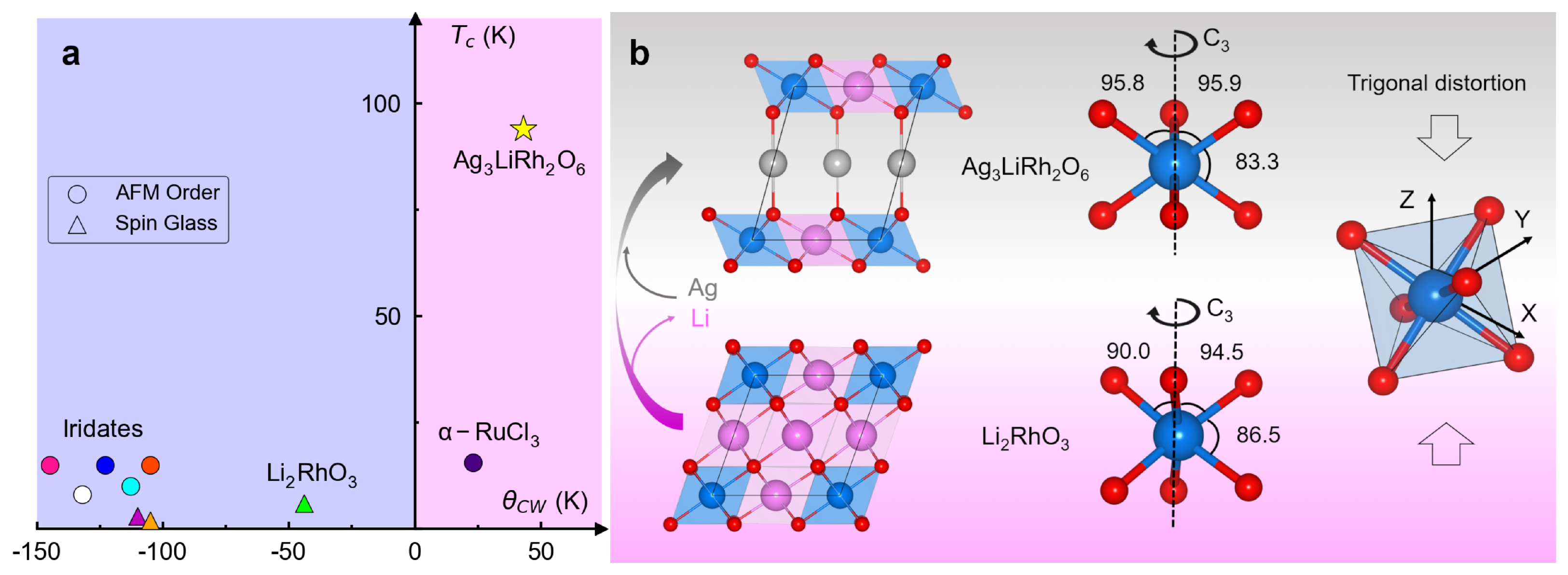}
 \caption{\label{fig:PD}
 \textbf{Phase diagram.}
 (\textbf{a}) Critical temperature ($T_c$) plotted against the Curie-Weiss temperature ($\Theta_\text{CW}^{\text{avg}}$) using the data in Table~S1 for polycrystalline 2D Kitaev materials.
 Circles and triangles represent AFM and spin-glass transitions, respectively.
The iridate materials are (from left to right) \CLIO, \ALIO, \NIO, \CNIO, \CIO, \HLIO, and \LIO.
 (\textbf{b}) Structural relationship between the first and second-generation Kitaev systems, \LRO\ and \ALRO, with enhanced trigonal distortion in the latter, as evidenced by the change of bond angles after cation exchange.
 }
\end{figure*}

Our experiment was motivated by a survey of the average Curie-Weiss temperature ($\Theta_\text{CW}^{\text{avg}}$) and the antiferromagnetic (AFM) or spin-glass transition temperatures ($T_\text{N}/T_\text{g}$) of the two-dimensional (2D) iridium, rhodium, and ruthenium-based Kitaev materials (Fig.~\ref{fig:PD}a and supplementary Table~S1).   
These compounds can be categorized into two groups.
The first-generation Kitaev magnets include \LIO, \NIO, \LRO, and \RCL, synthesized by conventional solid-state methods~\cite{singh_relevance_2012,mehlawat_heat_2017,singh_antiferromagnetic_2010,plumb_rucl3_2014,kobayashi_moessbauer_1992,koitzsch_j_mathrmeff_2016,banerjee_proximate_2016,mazin_origin_2013}.
The second-generation materials, such as \HLIO, \CNIO, and \ALIO, have been synthesized recently by exchanging the interlayer alkali (Li$^+$ and Na$^+$) in the first-generation compounds with H$^+$ , Cu$^+$, and Ag$^+$ using topochemical reactions~\cite{roudebush_iridium_2016,abramchuk_cu2iro3:_2017,bahrami_effect_2021,kenney_coexistence_2019,takahashi_spin_2019,bahrami_thermodynamic_2019,kitagawa_spinorbital-entangled_2018}.
Both the first and second-generation iridates appear in the same region of the phase diagram in Fig.~\ref{fig:PD}a.  
The $4d$ systems, namely \LRO\ and \RCL, appear to be shifted horizontally but not vertically from the iridate block.
Despite theoretical predictions of diverse magnetic phases~\cite{rusnacko_kitaev-like_2019,rau_generic_2014}, it seems all 2D Kitaev materials studied so far aggregate in the same region of the phase diagram with $T_\text{N}\le 15$~K and a $J_{\text{eff}}=\frac{1}{2}$ state. 
This observation prompted us to experimentally investigate the possibility of tuning the local spin-orbital state and the magnetic ground state in the same material family.

We focused on rhodate ($4d$) systems where the SOC is weaker than in the iridate ($5d$) systems, and $\Delta_\text{T}$ has a better chance to compete with $\lambda_ \text{SOC}$.
Evidence of such competition can be found in earlier density functional theory (DFT) studies of the honeycomb rhodates, where a high sensitivity of the magnetic ground state to structural parameters has been reported~\cite{katukuri_strong_2015,mazin_origin_2013}. 
To enhance $\Delta_\text{T}$, we replaced the Li atoms between the layers of \LRO\ with Ag atoms, and synthesized \ALRO\ topochemically (Fig.~\ref{fig:PD}b).
The change of interlayer bonds leads to a trigonal compression along the local $C_3$ axis (Fig.~\ref{fig:PD}b).
Using crystallographic refinement (supplementary Fig.~S1 and Tables~S2 and S3), we determined the bond angles within the local octahedral ($O_h$) environments of both compounds and quantified the trigonal distortion by calculating the bond angle variance~\cite{haraguchi_strong_2020} $\sigma=\sqrt{\sum_{i=1}^{12} (\theta - \theta_0)^2 / (m-1)}$, where $m=12$ and $\theta_0=90^\circ$.
In an ideal octahedron, $\sigma=0$.
In \ALRO, we found $\sigma=6.1(1)^\circ$, nearly twice the $\sigma=3.1(1)^\circ$ in \LRO.
It has been noted in earlier theoretical works~\cite{khaliullin_orbital_2005,bhattacharjee_spinorbital_2012} that a trigonal distortion can reconstruct the spin-orbital states and lead to new magnetic regimes; however, it has also been noted that such a regime may not be accessible in iridate materials due to the overwhelmingly strong SOC.
As shown in Fig.~\ref{fig:PD}a, we induced such a change of regime between \LRO\ and \ALRO\ using chemical pressure.

\section*{Results}

\subsection*{Magnetic Properties}
\begin{figure*}[ht]
 \centering
 \includegraphics[width=\textwidth]{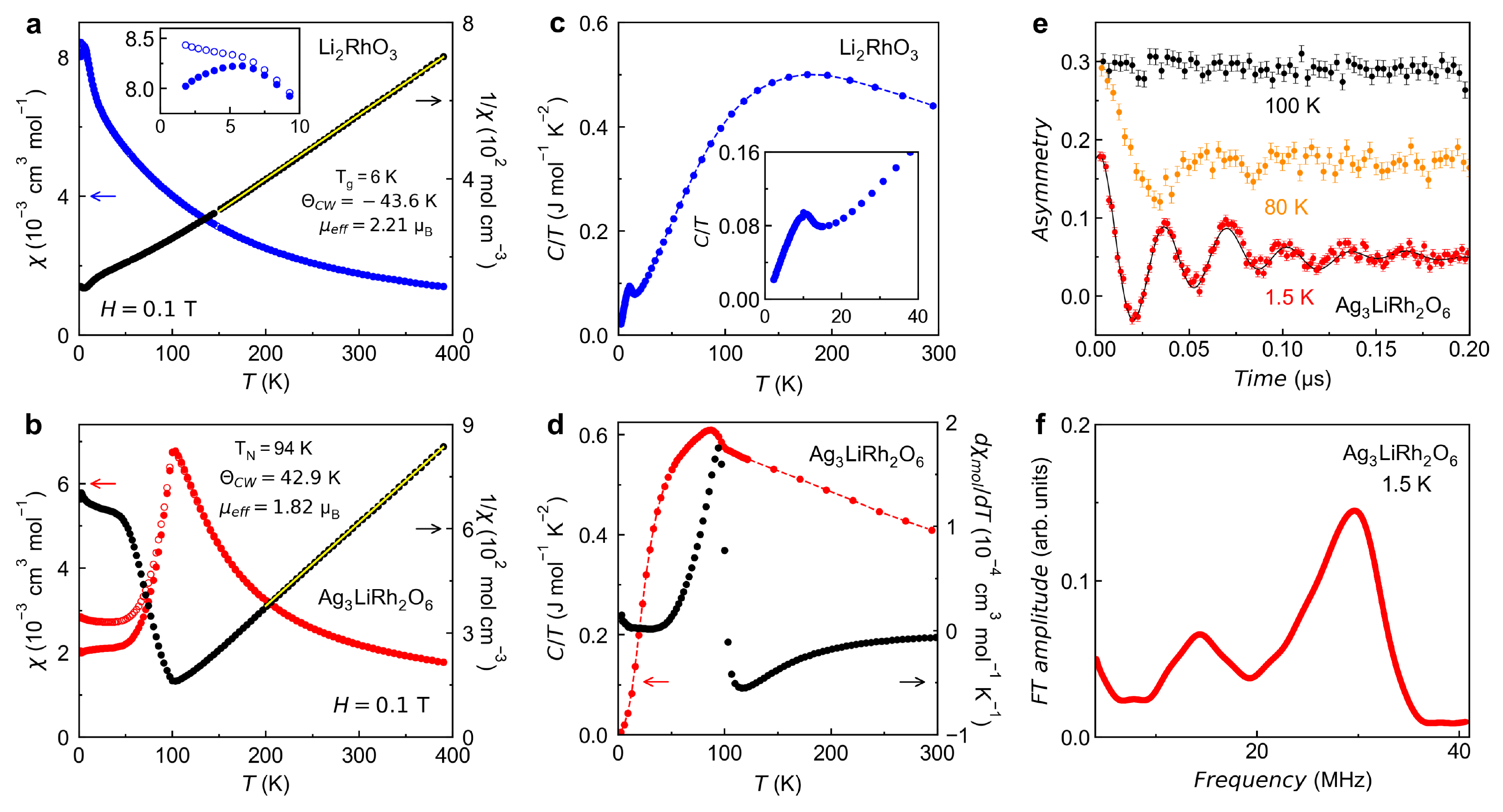}
 \caption{\label{fig:MTH}
 \textbf{Magnetic characterization.}
 Magnetic susceptibility plotted as a function of temperature and Curie-Weiss analysis presented in (\textbf{a}) \LRO\ (blue) and (\textbf{b}) \ALRO\ (red).
The ZFC and FC data are shown as full and empty symbols, respectively.
 Heat capacity as a function of temperature in (\textbf{c}) \LRO\ and (\textbf{d}) \ALRO.
The black circles in (\textbf{d}) show the derivative of magnetic susceptibility with respect to temperature.
 (\textbf{e}) $\mu$SR asymmetry plotted as a function of time in \ALRO. 
 For clarity, the curves at 100 and 80~K are offset with respect to the 1.5~K spectrum. 
 The solid line is a fit to a Bessel function (see the supplementary materials for details).
 (\textbf{f}) Fourier transform of the  $\mu$SR spectrum at 1.5~K showing two frequency components.
 }
\end{figure*}
A small peak at $T_\mathrm{g}=6.0(5)$~K in \LRO\ with a splitting between the zero-field-cooled (ZFC) and field-cooled (FC) susceptibility data ($\chi(T)$ in Fig.~\ref{fig:MTH}a) confirms the spin-glass transition as reported in earlier works~\cite{luo_li_2rho_3_2013,mazin_origin_2013,khuntia_local_2017}.
In stark contrast, \ALRO\ exhibits a robust AFM order with a pronounced peak in $\chi(T)$ and without ZFC/FC splitting (Fig.~\ref{fig:MTH}b).
The small difference between the ZFC and FC curves at low temperatures is due to a small amount of stacking faults, which are carefully analyzed in supplementary Fig.~S2.
A Curie-Weiss analysis in Fig.~\ref{fig:MTH}b yields an effective moment of $1.82~\mu_B$ and a $\Theta_\text{CW}^{\text{avg}}=42.9$~K, consistent with a prior report~\cite{todorova_agrho2_2011}.
A positive $\Theta_\text{CW}^{\text{avg}}$ despite an AFM order suggests that $\chi(T)$ must be highly anisotropic, which is the case in materials with A-type or C-type AFM order.
For example, Na$_3$Ni$_2$BiO$_6$ has a C-type AFM order (AFM intralayer and ferromagnetic (FM) interlayer) with $T_\text{N}=10.4$~K and $\Theta_\text{CW}^{\text{avg}}=13.3$~K~\cite{seibel_structure_2013}.

Both the spin-glass transition in \LRO\ and the AFM transition in \ALRO\ are marked by peaks in the heat capacity in Fig.~\ref{fig:MTH}c,d.
The heat capacity peak of \ALRO\ is visible despite the large phonon background at high temperatures, confirming a robust AFM order. 
We report $T_\text{N}=94(3)$~K using the peak in $d\chi/dT$, which is close to the peak in the heat capacity (Fig.~\ref{fig:MTH}d).
$T_\text{N}$ in \ALRO\ is nearly an order of magnitude larger than the transition temperature in any other 2D Kitaev material to date.

To obtain information about the local field within the magnetically ordered state of \ALRO\ we turned to $\mu$SR experiments.
In Fig.~\ref{fig:MTH}e, the time-dependent $\mu$SR asymmetry curves in zero applied magnetic field show the appearance of spontaneous oscillations below 100~K, confirming the long-range magnetic order.
The asymmetry spectrum at 1.5 K fits to a modified zeroth-order Bessel function~\cite{amato_understanding_2014}, with the form of the fitting function indicating non-collinear incommensurate magnetic ordering (see supplementary materials for details).
As shown in Fig.~\ref{fig:MTH}f, the Fourier transform of the 1.5~K spectrum shows two peaks at 12 and 31~MHz, which we have modeled using a two-component expression.
Each component has a distribution of local fields between a $B_{min}$ and $B_{max}$ indicating incommensurate ordering (supplementary Table~S4).
The center of distribution $(B_{min}+B_{max})/2$ is shifted from zero, indicating a non-collinear order~\cite{amato_understanding_2014}.

The dominant frequency of 31~MHz in Fig.~\ref{fig:MTH}f corresponds to a maximum internal field of 0.231~T at the muon stopping site (using $\omega=\gamma_\mu B$ with the muon gyromagnetic ratio $\gamma_\mu=851.6$ Mrad\,s$^{-1}$T$^{-1}$), which is an order of magnitude larger than the internal field of 0.015~T extracted from $\mu$SR in \LRO~\cite{khuntia_local_2017}.
The dramatic change of magnetism between \LRO\ and \ALRO\ in response to mild trigonal distortion, with an order-of-magnitude increase in both $T_\text{N}$ and internal field, implies a novel underlying interaction in the ground state.

\subsection*{Theoretical wave functions}
\begin{figure*}[ht]
 \centering
 \includegraphics[width=\textwidth]{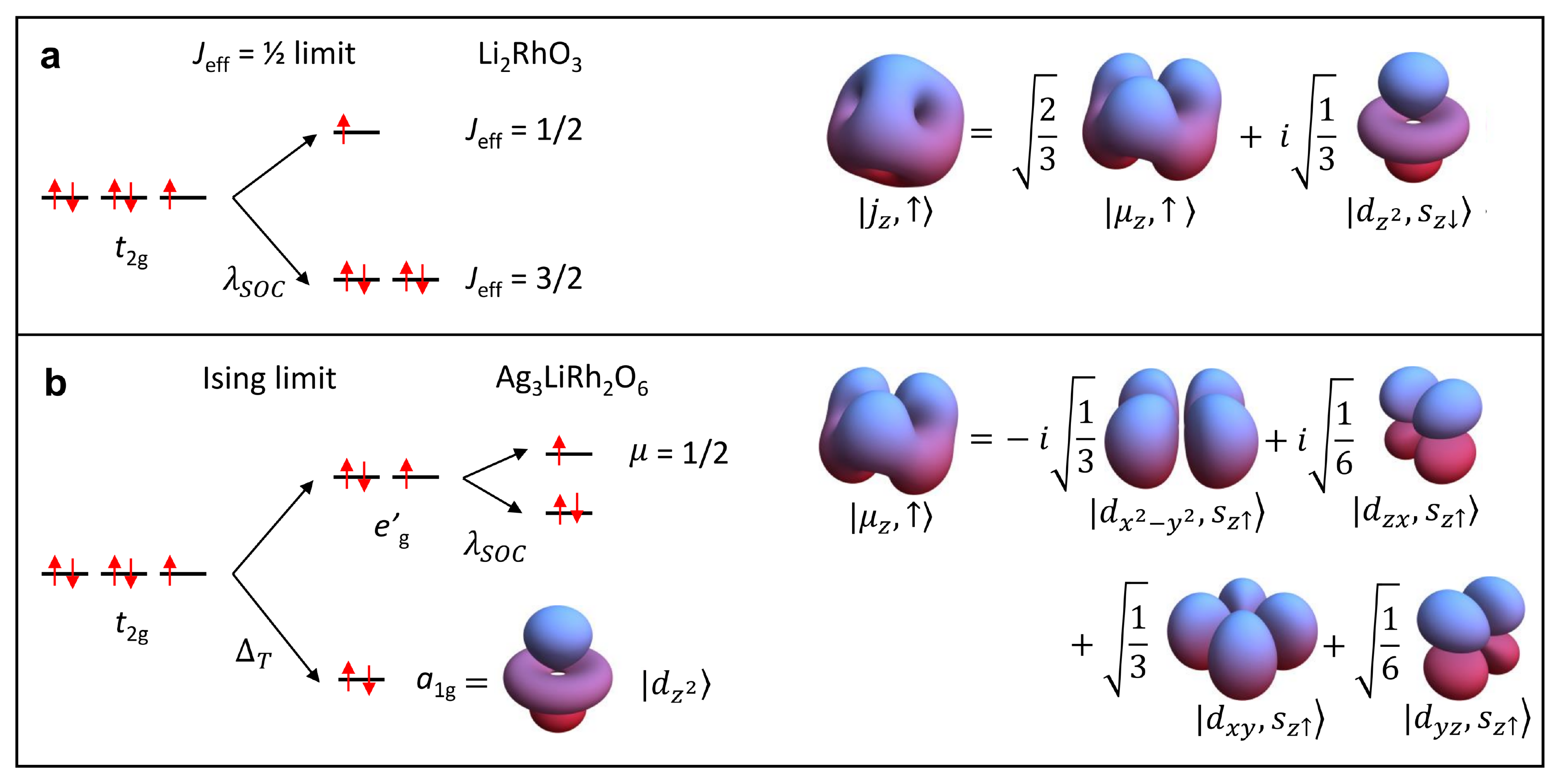}
 \caption{\label{fig:THEORY}
 \textbf{Wave functions.}
 (\textbf{a}) The $J_{\text{eff}}=1/2$ limit, realized in \LRO, where $\lambda_{\text{SOC}}\gg \Delta_{T}$.
 The probability density is visualized for the isospin-up wave function.
 (\textbf{b}) The Ising limit, realized in \ALRO, where $\Delta_T\gg \lambda_{\text{SOC}}$.
 The probability density is visualized for the spin-up wave function.
 Notice the cubic and trigonal symmetries of the $J_z$ and $\mu_z$ orbitals, respectively.
 }
\end{figure*}
The drastic change of magnetic behavior between \LRO\ and \ALRO\ originates from a fundamental change of the spin-orbital quantum state (Fig.~\ref{fig:THEORY}). 
Both \LRO\ and \ALRO\ have Rh$^{4+}$ in the $4d^5$ configuration, corresponding to one hole in the \tg\ manifold. 
Assuming that both compounds are in the Mott insulating regime (supplementary Fig.~S3), their low-energy physics should be described by a Kramers doublet per Rh$^{4+}$ ion, i.e. they are effective spin-$\frac{1}{2}$ systems. 
However, the nature of the Kramers doublet may be significantly different depending on the interplay between $\lambda_{\text{SOC}}$ and $\Delta_\text{T}$. 
We illustrate this by considering two limits: the $J_{\text{eff}}=1/2$ limit for $\lambda_{\text{SOC}}\gg\Delta_\text{T}$ relevant to \LRO\ (Fig.~\ref{fig:THEORY}a) and the Ising limit for $\Delta_\text{T}\gg\lambda_{\text{SOC}}$ relevant to \ALRO\ (Fig.~\ref{fig:THEORY}b). 
The wave functions of the low-energy Kramers doublet can be found in both limits using perturbation theory. 
The Kramers doublet in the $J_{\text{eff}}=1/2$ limit ($\lambda_{\text{SOC}}\gg\Delta_\text{T}$) comprises the following two states (Fig.~\ref{fig:THEORY}a):
\begin{equation}
|j_{z,\uparrow}\rangle=\sqrt{\frac{2}{3}}|\mu_{z},\uparrow\rangle+i\sqrt{\frac{1}{3}}|d_{z^2},s_{z\downarrow}\rangle,\quad|j_{z,\downarrow}\rangle=\sqrt{\frac{2}{3}}|\mu_{z},\downarrow\rangle+i\sqrt{\frac{1}{3}}|d_{z^2},s_{z\uparrow}\rangle \label{eq:Jeff_limit},
\end{equation}
where $\{|\mu_{z,\uparrow}\rangle,|\mu_{z,\downarrow}\rangle\}$ are defined in Eq.~\ref{eq:Ising_limit} below. 
The $J_{\text{eff}}=1/2$ limit has been discussed extensively in the literature, and sizable Kitaev interactions have been proposed for materials in this limit such as the honeycomb iridates~\cite{takagi_concept_2019,knolle_field_2019,chaloupka_kitaev-heisenberg_2010,jackeli_mott_2009}, \RCL~\cite{koitzsch_j_mathrmeff_2016}, and \LRO~\cite{luo_li_2rho_3_2013,khuntia_local_2017,katukuri_strong_2015}.
The only difference between Eq.~\ref{eq:Jeff_limit} and prior works~\cite{jackeli_mott_2009} is that we choose the $z$-axis to be normal to the triangular face of the octahedron (Fig.~\ref{fig:PD}b) instead of pointing at the apical oxygens.

In the Ising-limit ($\Delta_\text{T}\gg\lambda_{\text{SOC}}$), the trigonal distortion leads to new Kramers doublet states (Fig.~\ref{fig:THEORY}b).
\begin{align}
|\mu_{z,\uparrow}\rangle &= -i\sqrt{\frac{1}{3}}|d_{x^2-y^2},s_{z\uparrow}\rangle+i\sqrt{\frac{1}{6}}|d_{zx},s_{z\uparrow}\rangle + \sqrt{\frac{1}{3}}|d_{xy},s_{z\uparrow}\rangle+\sqrt{\frac{1}{6}}|d_{yz},s_{z\uparrow}\rangle, \notag\\
|\mu_{z,\downarrow}\rangle &= i\sqrt{\frac{1}{3}}|d_{x^2-y^2},s_{z\downarrow}\rangle-i\sqrt{\frac{1}{6}}|d_{zx},s_{z\downarrow}\rangle + \sqrt{\frac{1}{3}}|d_{xy},s_{z\downarrow}\rangle+\sqrt{\frac{1}{6}}|d_{yz},s_{z\downarrow}\rangle.\label{eq:Ising_limit}
\end{align}
Note that the states $\{|j_{z,\uparrow}\rangle,|j_{z,\downarrow}\rangle\}$ are \emph{not} orthogonal to states $\{|\mu_{z,\uparrow}\rangle,|\mu_{z,\downarrow}\rangle\}$, despite being in opposite limits.

The trigonal splitting energy scale $\Delta_\text{T}$ is known to split the six-fold degenerate \tg\ levels (including spin degrees of freedom) into a two-fold $a_\text{1g}$ manifold and a four-fold $e'_\text{g}$ manifold (Fig.~\ref{fig:THEORY}b)~\cite{joy_magnetism_1992}. 
Choosing $\hat x$ and $\hat y$ directions pointing towards oxygen atoms as shown in Fig.~\ref{fig:PD}b, the orbital wave functions are found to be $|d_{z^2}\rangle$ for $a_{1g}$ (Fig.~\ref{fig:THEORY}b), and $\{|\tau_{z,\uparrow}\rangle,|\tau_{z,\downarrow}\rangle\}$ for $e'_\text{g}$:
\begin{equation}
   |\tau_{z,\uparrow}\rangle\equiv\sqrt{\frac{2}{3}}|d_{x^2-y^2}\rangle-\sqrt{\frac{1}{3}}|d_{zx}\rangle,\quad |\tau_{z,\downarrow}\rangle\equiv\sqrt{\frac{2}{3}}|d_{xy}\rangle+\sqrt{\frac{1}{3}}|d_{yz}\rangle,\label{eq:egp}
\end{equation}
In the materials under consideration, the trigonal distortion is a compression along the $\hat z$-axis (Fig.~\ref{fig:PD}b) that lowers the energy of the $a_{1g}$ level (Fig.~\ref{fig:THEORY}b).
Thus, for the $4d^5$ configuration, one should focus on the four-fold $e'_\text{g}$ manifold.
Unlike in the \eg\ manifold, the spin-orbit coupling is not completely quenched in the $e'_\text{g}$ manifold. 
We show, in the supplementary materials, that the $d$-orbital angular momentum operator $\vec L$, after projection into the $e'_{g}$ manifold, becomes:
\begin{equation}
    L_x\rightarrow 0,\quad L_y\rightarrow 0,\quad L_z\rightarrow \tau_y,
\end{equation}
where $\tau_{x,y,z}$ are the pseudospin Pauli matrices.
We therefore have, in the $e'_\text{g}$ manifold
\begin{equation}
\lambda_{\text{SOC}}~\vec L\cdot\vec s\rightarrow \lambda_{\text{SOC}}~L_z s_z= \lambda_{\text{SOC}}~\tau_y s_z
\end{equation}
Namely, $\lambda_{\text{SOC}}$ further splits the $e'_\text{g}$ manifold into two Kramers doublets: $\tau_y$ anti-aligned with $s_z$, or $\tau_y$ aligned with $s_z$. 
The former doublet has a lower energy, so the latter doublet is half-filled in the $4d^5$ configuration. 
Finally, the low-energy effective spin-$\frac{1}{2}$ states in the Ising-limit are
\begin{equation}
    |\mu_{z,\uparrow}\rangle=|\tau_y=+1,s_{z,\uparrow}\rangle,\quad |\mu_{z,\downarrow}\rangle=|\tau_y=-1,s_{z,\downarrow}\rangle
\end{equation}
which are nothing but the states written in Eq.~\ref{eq:Ising_limit} and illustrated in Fig.~\ref{fig:THEORY}b.

The exchange couplings for the effective $\mu$-spins are expected to have the Ising anisotropy (easy-axis along the $\hat z$-direction). 
To understand its origin, one may consider exchange interactions like $J\vec S_i\cdot\vec S_j$ between two Rh sites $i,j$ in the absence of the spin-orbit interaction. 
After $\lambda_{\text{SOC}}$ is turned on, the $J\vec S_i\cdot\vec S_j$ needs to be projected onto the Kramers doublet $\{|\mu_{z,\uparrow}\rangle, |\mu_{z,\downarrow}\rangle\}$ at low energies. 
Only the term $J S_{i,z}S_{j,z}$ survives after the projection. 
In addition, the g-factor of the effective $\mu$-spins in a magnetic field is also expected to be significantly anisotropic. 
For example, the effective $\mu$-spins do not couple with a magnetic field along the $\hat x$ (or $\hat y$) axis in a linear fashion in this limit.
A direct measurement of the magnetic response with respect to the field direction is not possible at this stage because single crystals of \ALRO\ are not available.
However, indirect evidence of such anisotropic interactions may be the positive Curie-Weiss temperature in polycrystalline samples of \ALRO\ (Fig~\ref{fig:MTH}b) that indicates FM interactions despite the AFM ordering.
Such a behavior has been reported in Na$_3$Ni$_2$BiO$_6$ and attributed to a C-type AFM order where the coupling within the layers is AFM and between the layers is FM~\cite{seibel_structure_2013}.

\subsection*{Spectroscopic evidence}
\begin{figure*}[ht]
\centering
\includegraphics[width=\textwidth]{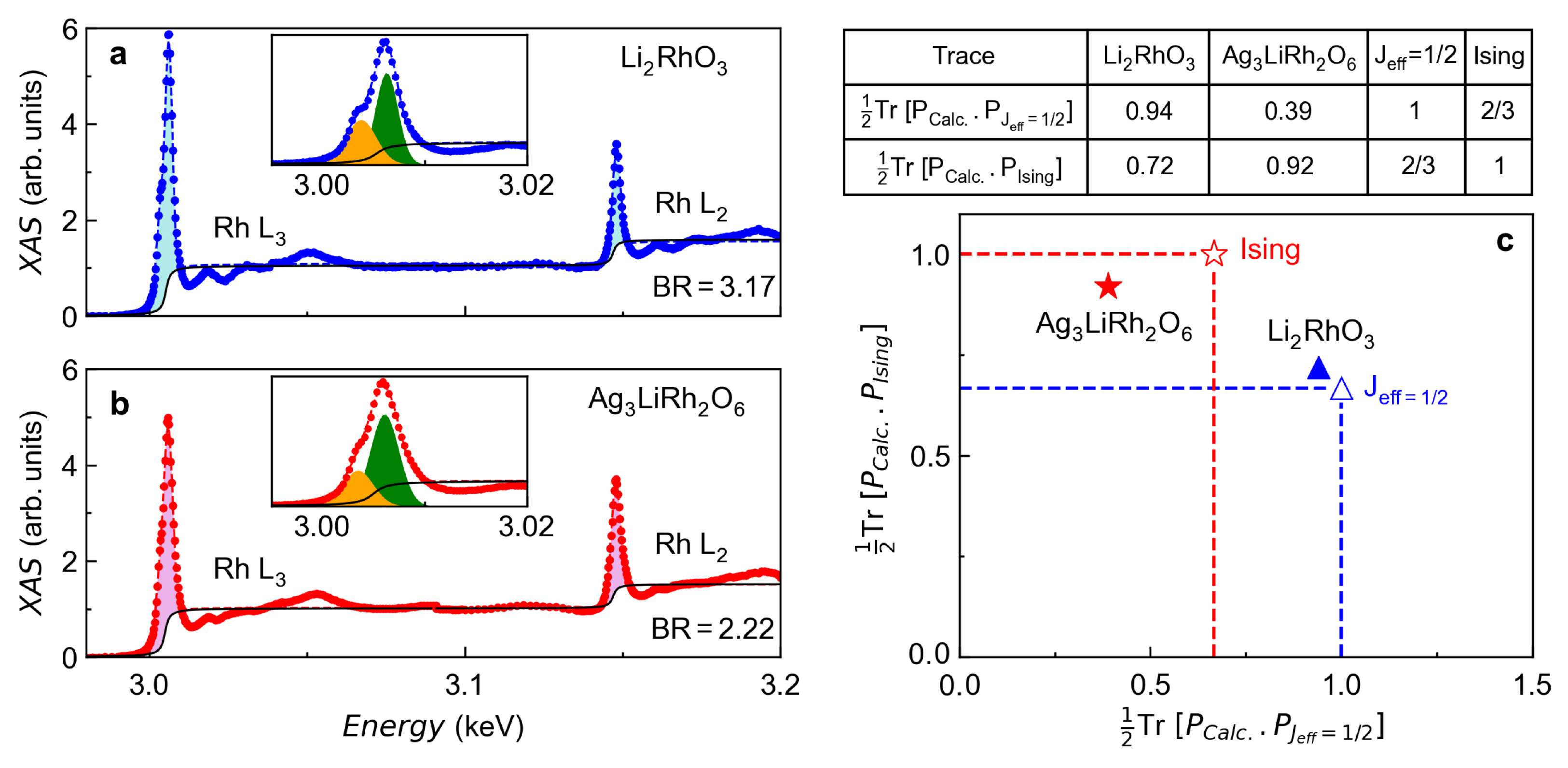}
\caption{\label{fig:XAS}
\textbf{X-ray absorption spectroscopy (XAS).}
(\textbf{a}) XAS data from Rh $L_{2,3}$ edges of \LRO. 
The data were modeled with a step and two Gaussian functions for the $L_3$ edge (inset), and one Gaussian function for the $L_2$ edge.
(\textbf{b}) Similar data and fits for the Rh $L_{2,3}$ edges of \ALRO.
(\textbf{c}) Theoretically calculated traces of projector products are tabulated and plotted for both the ideal limits (empty symbols) and real materials (full symbols).
}
\end{figure*}
We provide spectroscopic confirmation of the above picture by measuring the branching ratio using x-ray absorption spectroscopy (XAS).
Figure~\ref{fig:XAS}a shows the XAS data from a \LRO\ sample with the Rh $L_3$ and $L_2$ edges near 3.00 and 3.15~keV, respectively.
The branching ratio, $BR=I(L_3)/I(L_2)=3.17(1)$, is evaluated by dividing the shaded areas under the $L_3$ and $L_2$ peaks in Fig.~\ref{fig:XAS}a.
A similar analysis in \ALRO\ yields $BR=2.22(1)$ (Fig.~\ref{fig:XAS}b).
The branching ratio is related to the SOC through $BR=(2+r)/(1-r)$, where $r=\langle \mathbf{L} \cdot \mathbf{S} \rangle/n_h$ with $n_h$ being the number of holes in the $4d$ shell~\cite{van_der_laan_local_1988,laguna-marco_orbital_2010}. 
Using $n_h=5$ for Rh$^{4+}$, we obtain $\langle \mathbf{L} \cdot \mathbf{S} \rangle=1.40$ in \LRO~and 0.34 in \ALRO.
This is consistent with the above theoretical picture based on $\lambda_{\text{SOC}}\gg\Delta_\text{T}$ and the \jeff\ limit in \LRO\ compared to $\Delta_\text{T}\gg \lambda_{\text{SOC}}$ and the Ising limit in \ALRO.

Note that the spectroscopic value of $\langle \mathbf{L} \cdot \mathbf{S} \rangle$ is small but non-vanishing in \ALRO. 
The fine structure of Rh $L$ edge with a shoulder near the $L_3$ peak (inset of Fig.~\ref{fig:XAS}b) that is absent in the $L_2$ peak confirms a finite SOC in \ALRO~\cite{de_groot_differences_1994}.
As illustrated in Fig.~\ref{fig:THEORY}b, a weak SOC is necessary to split the \eeg\ levels.
The fine structure of the Rh $L_3$ edge can be fitted to two Gaussian curves in both \LRO\ and \ALRO\ (insets of Fig.~\ref{fig:XAS}a,b).
A higher ratio between the two Gaussian areas in \ALRO\ (2.42) than in \LRO\ (1.53) is consistent with a weaker SOC in the former. 
Supporting information about the Ag $L$ edge is provided in Fig.~S4 to confirm the Ag$^+$ oxidation state~\cite{kolobov_thermal_2004,behrens_electronic_1999}.

\section*{Discussion}
We have demonstrated that a competition between SOC and trigonal distortion could tune a honeycomb structure between the Kitaev ($J_\text{eff}=\frac{1}{2}$) and Ising ($\mu=\frac{1}{2}$) limits.
Our magnetization, $\mu$SR, and XAS data suggest that \LRO\ is closer to the \jeff\ limit, whereas \ALRO\ is closer to the Ising limit.
We calculated the ideal wave functions and visualized them in Fig.~\ref{fig:THEORY}.
However, a realistic material is generally situated on a spectrum between these two ideal limits. 
For example, minor lattice distortions (e.g. monoclinic) can further break the trigonal point group symmetry and perturb the ideal wave function.

To make our theoretical discussion more realistic, we calculated the band structure of \LRO\ and \ALRO\ from first principles and obtained a real-space tight-binding model for each compound.

Details of the electronic structure calculations in the presence of Hubbard-$U$, SOC, and zigzag magnetic ordering are presented in supplementary Figs.~S5 and S6, and Table~S5.  
The full orbital content of the energy eigenstates was characterized using a combination of Quantum-Espresso and Wannier90 softwares~\cite{giannozzi2009quantum,giannozzi2017advanced,Pizzi2020}.

This allowed us to quantitatively investigate the regimes being realized in \LRO\ and \ALRO. 
Specifically, given a Kramers doublet $|\psi_1\rangle$ and $|\psi_2\rangle$, we defined the projectors:
\begin{equation}
  P_\psi\equiv |\psi_1\rangle\langle\psi_1|+|\psi_2\rangle\langle\psi_2|
\end{equation}
For example, $P_{J_{\text{eff}}=1/2}$ and $P_{\text{Ising}}$ are projectors defined using the Kramers doublets in the $J_{\text{eff}}=1/2$ limit (Eq.~\ref{eq:Jeff_limit}) and the Ising limit (Eq.~\ref{eq:Ising_limit}), respectively. 
We then compute the traces $\frac{1}{2}\mathrm{Tr}[P_{calc.}\cdot P_{J_{\text{eff}}=1/2}]$ and $\frac{1}{2}\mathrm{Tr}[P_{calc.}\cdot P_{\text{Ising}}]$. 
The results are tabulated and visualized in Fig.~\ref{fig:XAS}c.
These traces would be unity if the calculated system was in the ideal $J_\text{eff}$ or Ising limit. 
Figure~\ref{fig:XAS}c clearly locates \LRO\ and \ALRO\ in the vicinity of the $J_{\text{eff}}=1/2$ and Ising ($\mu=1/2$) limits, respectively.

Our combined experimental and theoretical results show how to change the fabric of spin-orbit coupled states and dramatically change the magnetic behavior of the Kitaev materials.
Despite theoretical proposals for a diverse global phase diagram, the current Kitaev systems are all in the $J_\text{eff}$ limit~\cite{sears_ferromagnetic_2020,knolle_field_2019,chaloupka_kitaev-heisenberg_2010,jackeli_mott_2009,kim_novel_2008,gretarsson_crystal-field_2013,clancy_pressure-driven_2018,plumb_rucl3_2014,koitzsch_j_mathrmeff_2016,takagi_concept_2019}.
Finding an outlier, such as \ALRO, in the phase diagram (Fig.~\ref{fig:PD}a) provides the first glimpse at the diversity of magnetic phases that can be engineered using topochemical methods.
Specifically, the interplay between the Kitaev and Ising limits will be a fruitful venue to search for novel non-collinear magnetic orders beyond the familiar Kitaev-Heisenberg paradigm.

\section*{Materials and Methods}

\subsection*{Material Synthesis.}
Similar to other second-generation Kitaev magnets, \ALRO\ is a metastable compound.
It is synthesized through a topotactic cation-exchange reaction under mild conditions from the first-generation parent compound \LRO.
\begin{equation}
\label{eq:reaction}
\mathrm{2Li_2RhO_3 + 3AgNO_3 \longrightarrow Ag_3LiRh_2O_6 + 3LiNO_3}
\end{equation}
\LRO\ was synthesized following prior published works~\cite{luo_li_2rho_3_2013,khuntia_local_2017}.

\subsection*{Characterizations.} Powder x-ray diffraction (PXRD) was performed using a Bruker D8 ECO instrument in the Bragg-Brentano geometry, using a copper source (Cu-K$_{\alpha}$) and a LYNXEYE XE 1D energy dispersive detector.
The FullProf suite was used for the Rietveld analysis~\cite{rodriguez-carvajal_recent_1993}.
Peak shapes were modeled with the Thompson-Cox-Hastings pseudo-Voigt profile convoluted with axial divergence asymmetry.
Magnetization was measured using a Quantum Design MPMS3 with the powder sample mounted on a low-background brass holder.
Both the electrical resistivity (four-probe technique) and heat capacity (relaxation-time method) were measured on a pressed pellet using the Quantum Design PPMS Dynacool.
Electron diffraction (ED), high-angle annular dark-field scanning TEM (HAADF-STEM), and annular bright-field scanning TEM (ABF-STEM) were performed using an aberration double-corrected JEM ARM200F microscope operated at 200~kV and equipped with a CENTURIO EDX detector, Orius Gatan CCD camera and GIF Quantum spectrometer~\cite{houska_leastsquares_1981,abramchuk_crystal_2018,krivanek_atom-by-atom_2010}.
The $\mu$SR measurements were performed in a continuous flow $^4$He evaporation cryostat ($T>1.5$~K) at the General Purpose Surface-Muon Instrument (GPS)~\cite{amato_new_2017} at the Paul Scherrer Institute (PSI), and the data was analyzed using the Musrfit program~\cite{suter_musrfit_2012}. 
A pressed disk of \ALRO\ with diameter 12~mm and thickness 1.2~mm was wrapped in 25~$\mu$m silver foil and suspended in the muon beam to minimize the contribution from muons implanted in a sample holder or in the cryostat walls.

\subsection*{First principle calculations.}
The electronic structures were computed using the open-source code QUANTUM ESPRESSO~\cite{giannozzi2009quantum,giannozzi2017advanced} with the experimental crystallographic information as the input. 
The calculation included spin-orbit coupling and zigzag magnetic ordering for both compounds, and the fully gapped states were achieved using a DFT+U method~\cite{cococcioni_linear_2005}. 
To stabilize the non-collinear magnetic calculation, we used the norm-conserving (NC) pseudopotentials from PseudoDojo~\cite{van_setten_pseudodojo_2018}. 
For convergence reasons we implemented the Perdew-Zunger (PZ) functional in the calculation of \ALRO, while leaving the default Perdew-Burke-Ernzerhof (PBE) functional for \LRO. 
To compare with the previous report of the insulating states in \LRO\cite{mazin_origin_2013}, we fixed the Hund's coupling to $J=0.7$~eV and tuned the Hubbard-$U$ from 1 to 4~eV. 
Our results were consistent with the prior work.
The real-space tight-binding functions (involving the Rh-$4d$ and O-$2p$ orbitals as well as Ag-$4d$ orbitals for \ALRO) were derived from the band structure using maximally-localized Wannier states implemented by the Wannier90 software \cite{Pizzi2020}. 
From here, tight-binding models for a single $\mathrm{RhO_6}$ cluster were constructed based on the obtained real-space hopping parameters. 
The eigenstates of such a $\mathrm{RhO_6}$ cluster were used to compute $\langle \mathbf{L} \cdot \mathbf{S} \rangle$ and the traces in Fig.~\ref{fig:XAS}.

\subsection*{X-ray absorption spectroscopy.}
X-ray absorption near edge structure (XANES) data at Rh and Ag $L_{2,3}$ edges were collected at tender energy beamline 8-BM of the National Synchrotron Light Source II, and at beamline 4-ID-D of the Advanced Photon Source, respectively. 
The Rh $L_{2,3}$ data were collected in total electron yield (TEY) mode using powder samples in a helium gas environment. 
The Ag $L_{2,3}$ data were collected in partial fluorescence yield (PFY) mode with powder samples in vacuum. 
Silicon and nickel mirrors together with detuning of the second Si(111) monochromator crystal were used to reject high-energy harmonics. 
The PFY data were corrected for self-absorption~\cite{haskel}.


\bibliography{Bahrami_April152022}

\pagebreak

\noindent \textbf{Acknowledgements:}

The authors thank R.~Valenti and Y.~Li for fruitful discussions.
F.T. and F.B. acknowledge support from the National Science Foundation under Grant No. DMR-1708929.
Y.R. and X.H. acknowledge support from the National Science Foundation under Grant No. DMR-1712128.
This work is based in part on experiments performed at the Swiss Muon Source S$\mu$S, Paul Scherrer Institute, Villigen, Switzerland.
This research used 8-BM of the National Synchrotron Light Source II, a U.S. Department of Energy (DOE) Office of Science User Facility operated for the DOE Office of Science by Brookhaven National laboratory under Contract No. DE-SC0012704.
Work at the Advanced Photon Source was supported by the U.S. Department of Energy Office of Science, Office of Basic Energy Sciences, under Award No. DE-AC02-06CH11357.
\\
\noindent \textbf{Author Contributions:} F.B. synthesized the materials, performed magnetic and thermodynmic measurements, and analyzed the data.  O.I.L. performed TEM experiments. X.H. and Y.R. performed theoretical calculations. Y.D., G.F., and D.H. performed XAS experiments. C.W., H.L., and M.J.G. performed $\mu$SR experiments. F.T. conceptualized the project. All authors participated in writing.\\
\noindent \textbf{Competing Interests:} The authors declare no competing interests.\\

\end{document}